\documentclass[fleqn,usenatbib]{mnras}

\usepackage{newtxtext,newtxmath}

\usepackage[T1]{fontenc}

\DeclareRobustCommand{\VAN}[3]{#2}
\let\VANthebibliography\thebibliography
\def\thebibliography{\DeclareRobustCommand{\VAN}[3]{##3}\VANthebibliography}

\usepackage{graphicx}	
\usepackage{amsmath}	
\usepackage{subcaption}

\title[Short title, max. 45 characters]{Project Hephaistos – II. Dyson sphere candidates from Gaia DR3, 2MASS, and WISE}

\author[Suazo et al.]{Matías Suazo$^{1}$\thanks{E-mail: matias.suazo@physics.uu.se},
Erik Zackrisson$^{1,8}$,
Priyatam K. Mahto$^{2}$,
Fabian Lundell$^{1}$,
Carl Nettelblad$^{3}$,
Andreas J. Korn$^{1}$,
\newauthor
Jason T. Wright$^{4,5,6}$,
Suman Majumdar$^{2,7}$
\\
$^{1}$Observational Astrophysics, Department of Physics and Astronomy, Uppsala University, Box 516, SE-751 20 Uppsala, Sweden\\
$^{2}$Department of Astronomy, Astrophysics and Space Engineering, Indian Institute of Technology, Indore, India\\
$^{3}$Division of Scientific Computing, Department of Information Technology, Uppsala University, SE-751 05 Uppsala, Sweden\\
$^{4}$Department of Astronomy \& Astrophysics, The Pennsylvania State University, University Park, PA 16802, USA\\
$^{5}$Penn State Extraterrestrial Intelligence Center, The Pennsylvania State University, University Park, PA 16802, USA\\
$^{6}$Center for Exoplanets and Habitable Worlds, The Pennsylvania State University, University Park, PA 16802, USA\\
$^{7}$Department of Physics, Blackett Laboratory, Imperial College, London SW7 2AZ, UK\\
$^{8}$Swedish Collegium for Advanced Study, Linneanum, Thunbergsv\"a{}gen 2
SE-752 38 Uppsala, Sweden
}

\date{Accepted XXX. Received YYY; in original form ZZZ}

\pubyear{2015}

\begin{document}
\label{firstpage}
\pagerange{\pageref{firstpage}--\pageref{lastpage}}
\maketitle

\begin{abstract}
The search for extraterrestrial intelligence is currently being pursued using multiple techniques and in different wavelength bands. Dyson spheres, megastructures that could be constructed by advanced civilizations to harness the radiation energy of their host stars, represent a potential technosignature, that in principle may be hiding in public data already collected as part of large astronomical surveys. In this study, we present a comprehensive search for partial Dyson spheres by analyzing optical and infrared observations from Gaia, 2MASS, and WISE. We develop a pipeline that employs multiple filters to identify potential candidates and reject interlopers in a sample of five million objects, which incorporates a convolutional neural network to help identify confusion in WISE data. Finally, the pipeline identifies 7 candidates deserving of further analysis. All of these objects are M-dwarfs, for which astrophysical phenomena cannot easily account for the observed infrared excess emission.
\end{abstract}




\section{Introduction}

In the early 60s, \citet{dyson60} proposed an innovative methodology for searching for signs of extraterrestrial life. He presumed that highly advanced civilizations, in the pursuit of more energy resources, would construct an artificial, light-absorbing structure around their host star. This hypothetical structure, later referred to as a ``Dyson Sphere'', would allow them to harvest energy in the form of starlight. Starlight harvesting could, in principle, result in different observational signatures that may be detected using existing telescopes. These signatures include optical dimming of the host star due to direct obscuration, and waste-heat emission from the absorbing structure \citep[e.g.,][]{dyson60,wright16,Wright20}. Consequently, searching for anomalous infrared beacons in the sky has become an alternative to traditional communication-based searches for technologically advanced civilizations. One of the advantages of searches based on ``Dysonian'' signatures is that it does not rely on the willingness of other civilizations to contact us. 


Several observational projects have previously been conducted to detect individual Dyson spheres \citep[e.g.,][]{Slysh85,Jugaku91,Timofeev00,Jugaku04,Carrigan09,Zackrisson18} and for the large-scale use of similar technology at extragalactic distances \citep{Annis99,Wright14a,Wright14b,Griffith15,Zackrisson15,Garrett15,Lacki16,Olson17,Chen21}. However, none of these searches have revealed any strong candidates for Dysonian technology. 

Most search efforts have aimed for individual complete Dyson spheres, employing far-infrared photometry \citep[e.g.,][]{Slysh85, Jugaku91, Timofeev00, Carrigan09} from the Infrared Astronomical Satellite \citep[IRAS:][]{IRAS}, while a few considered partial Dyson spheres \citep[e.g.,][]{Jugaku04}. IRAS scanned the sky in the far infrared, providing data of $\approx 2.5\times 10^5$ point sources. However, nowadays, we rely on photometric surveys covering optical, near-infrared, and mid-infrared wavelengths that reach object counts of up to $\sim$$10^9$ targets and allow for larger search programs.

Within the context of {\it Project Hephaistos\footnote{\url{https://www.astro.uu.se/~ez/hephaistos/hephaistos.html}}}, in \citet{suazo22} we established upper limits on the prevalence of partial Dyson spheres in the Milky Way by analyzing the fraction of sources from Gaia DR2 and the Wide-field Infrared Survey Explorer (WISE) that exhibit infrared excess. In total, more than 10$^8$ stars were analyzed in that work. The exact upper limits on the fraction of stars that may host Dyson spheres reported by \citet{suazo22} are a function of distance, covering fraction and Dyson sphere temperature, but reach as low as $\sim$ 1 in 100,000 objects in the most constraining situation. However, the actual fraction is likely to be much lower (and possibly 0\%) since a number of other effects, such as dust emission and source blending, may also give rise to anomalous infrared fluxes. Note that the \citet{suazo22} upper limits are derived from color cuts rather than from fitting Dyson sphere models to the data, since the aim of that paper was not to discuss the nature of individual sources of excess infrared radiation.
 
 This second paper examines the Gaia DR3, 2MASS, and WISE photometry of $\sim$5 million sources to build a catalog of potential Dyson spheres. Here, we focus on the search for partial Dyson spheres, which partly obscure the starlight, which would still be detectable depending on the level of completion of the Dyson sphere. This structure would emit waste heat in the form of mid-infrared radiation that, in addition to the level of completion of the structure, would depend on its effective temperature. Gaia DR3 provides, unlike DR2, various astrophysical parameters derived from the low-resolution BP/RP spectra that can facilitate the rejection of  false positives in the search for Dyson spheres.
 
Gaia, 2MASS, and WISE all provide photometric data in the optical, near-infrared, and mid-infrared, respectively, but Gaia also provides parallax-based distances, which allow the spectral energy distributions of the targets to be converted to an absolute luminosity scale. The parallax data also make it possible to reject other point-like sources of strong mid-infrared radiation such as quasars, but do not rule out stars with a quasar in the background.

Since excess thermal emission at mid-infrared wavelengths represents the primary signature of Dyson spheres, searches for such objects naturally intersect with searches focused on mid-infrared excess sources in general. Excess emission in the infrared is a valuable tracer of the circumstellar dust that has been heated by the starlight and is reemitted at longer wavelengths. Circumstellar dust is present in structures such as young stars \citep[e.g.,][]{kennedy2012,kennedy2013,patel2014,cotten2016}. Many searches seeking infrared excess sources have encountered various difficulties when using WISE/AllWISE data, including flux overestimation for sources near the saturation limit \citep{cutri2013}, and the potential contamination from companion stars or background galaxies due to the large FWHM of the 12 and 22 $\mu m$ PSFs \citep[6.5'' and 12'' respectively; e.g.,][]{kennedy2012,theissen2017}. 



It has been proposed that Dyson spheres and similar radiation-harvesting megastructures could be constructed around a variety of stellar-mass objects, including white dwarfs \citep{Semiz15,zuckerman2022}, pulsars \citep{Osmanov16, Osmanov18} and black holes \citep{Hsiao21}. Here, we limit the discussion to Dyson spheres around main sequence stars. We additionally assume that feedback from Dyson spheres onto the host star may be neglected since this becomes relevant only when dealing with small, nearly-completed Dyson spheres or with highly internally reflective structures. \citep{Huston21}.

In Section~\ref{sec:methods}, we describe our overall search method. In Section~\ref{sec:results}, we present the most promising sources that emerged from our analysis, along with an examination of false positives encountered during the search. In Section~\ref{sec:discussion}, we discuss the likely nature of some of these Dyson sphere candidates and how future follow-up observations can help us disentangle their true nature. Section~\ref{sec:conclusions} summarizes our results.

\section{Methods}
\label{sec:methods}

This paper utilizes data from Gaia Data Release 3 \citep{gaia_space,gaiadr3}, 2MASS \citep{2mass}, and AllWISE \citep{cutri}. Gaia DR3 provides parallaxes and fluxes in three optical bands ($G_\mathrm{BP}$, $G$, $G_\mathrm{RP}$) in addition to various astrophysical parameters derived from the low-resolution BP/RP spectra. 2MASS provides near-infrared (NIR) fluxes in the J, H, and K$\rm _s$ bands, which corresponds to  1.2, 1.6, and 2.1 $\upmu$m, respectively, while WISE provides mid-infrared (MIR) fluxes at the W1, W2, W3, and W4 bands which corresponds to  3.4, 4.6, 12, and 22 $\upmu$m. The AllWISE program is an extension of the WISE program \citep{wright_wise} and combines data from different phases of the mission.

\begin{figure}
    \centering
    \includegraphics[width=0.9\columnwidth]{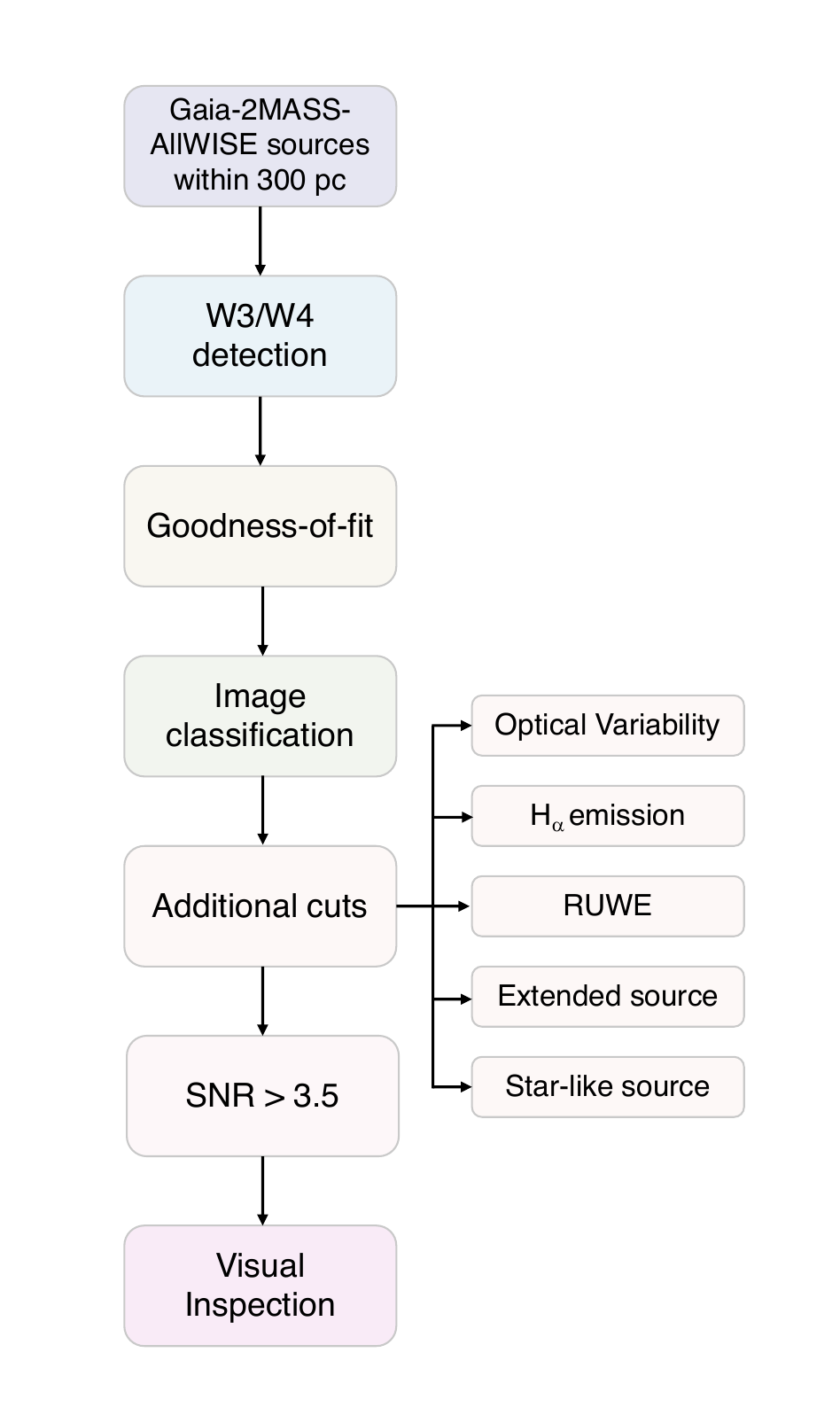}
    \caption{Flowchart illustrating our pipeline to find Dyson sphere candidates}
    \label{fig:pipeline}
\end{figure}

A specialized pipeline has been developed to identify potential Dyson sphere candidates, focusing on detecting sources that display anomalous infrared excesses that cannot be attributed to any known natural source of such radiation. It is essentially impossible to prove the existence of a Dyson spheres based on photometric data only, so this search can be considered a standard search for infrared excess sources biased towards excesses that are consistent with Dyson spheres based on their bright mid-infrared fluxes and our models of what the spectral energy distribution of Dyson spheres should look like. A simple schematic representation of this pipeline is illustrated in Figure~\ref{fig:pipeline}.


The pipeline for identifying Dyson sphere candidates involves several stages. We briefly describe each step:

\begin{itemize}
    \item Data Collection: We collect data from Gaia, 2MASS, and AllWISE for sources within 300 pc and detections in the 12 and 22 $\upmu$m bands (W3 and W4 WISE bands).
    \item Grid Search: A grid search method is employed to determine each star's best-fitting Dyson sphere model, utilizing the combined Gaia-2MASS-AllWISE photometry.
    \item Image Classification: To differentiate potential candidates located in nebular regions, a Convolutional Neural Network (CNN)-based algorithm is applied to WISE images to determine if our sources exhibit features associated with nebular regions. Young dust-obscured stars or stars otherwise associated with dusty nebulae appear as common false positives in our search. Therefore, only images lacking nebular features proceed to the 
    next step.
    \item Additional Analysis: This step involves utilizing several Gaia-WISE flags to assess whether the stars might exhibit an infrared excess of natural origin.
    \item Signal-to-noise ratio: Many sources with low signal-to-noise ratios (SNR in W3 and W4) slip through all the previous steps. Therefore we manually include this step where all sources with SNR lower than 3.5 in the W3 and W4 bands are rejected.
    \item Visual inspection: We visually inspect optical, near-, and mid-infrared images of all sources in order to reject problematic sources of mid-infrared radiation. Blends are the most typical confounder in this step.
\end{itemize}

These steps filter out sources that do not exhibit the desired characteristics of a Dyson sphere. Each step is explained in more detail in the following sections.


\subsection{Data Collection}
\label{sec:init}

We begin our search by taking a sample of stars from the Gaia DR3-2MASS-AllWISE catalog. The cross-matching between these catalogs was done by simultaneously using the \textit{allwise\_best\_neighbour}, \textit{tmass\_psc\_xsc\_best\_neighbour}, and \textit{tmass\_psc\_xsc\_join} catalogs provided by the Gaia consortium. Within this sample, our focus was on selecting stars located within a distance of 300 parsecs (pc) based on the geometric distance derived in the Early Data Release 3 (EDR3) \citep{Bailer-Jones21}. We opted to utilize EDR3 distances rather than Gaia DR3 distances, as the latter is derived from low-resolution BP/RP spectra and is therefore not available for most stars in the sample. 

Following the above mentioned criteria, our initial sample comprised approximately 5 million sources. Subsequently, we implemented an additional selection criterion, demanding detections in the 12 and 22 $\upmu$m bands (W3 and W4, respectively) from WISE. This choice was motivated by the fact that the expected infrared excess of Dyson spheres is particularly pronounced in these bands, given the range of temperature expected for Dyson spheres, as elaborated in Section~\ref{sec:model}. We additionally excluded sources that exhibited contamination according to the WISE contamination flag. As a result of this filtering step, our sample was downsized to approximately 320,000 stars.

\subsection{Theory and models}
\label{sec:model}

The next step in our pipeline corresponds to determining how well the photometry of the stars in the catalog resembles that of hypothetical main-sequence stars hosting Dyson spheres. This assessment requires understanding how the photometry of stars changes when surrounded by a Dyson sphere, which involves two effects: the obscuration of the star by the Dyson sphere and the re-emission of absorbed radiation by the structure at longer wavelengths.
To predict the observational characteristics of a composite system consisting of a star and a Dyson sphere (DS), we employ the model presented in \citet{suazo22}. This model incorporates the expected photometric fluxes of a DS into the photometry of observed main-sequence stars to simulate the combined system. In simple terms, the photometry of a star is modified according to the following equation:\begin{equation}
      M = -2.5\log(10^{-M_{\star}/2.5} + 10^{-M_{\rm DS}/2.5}),
      \label{eq:mag}
\end{equation} where $M_{\rm DS}$ represents the magnitude of the DS, and $M_{\star}$ corresponds to the magnitude of the star after it has been obscured by the Dyson sphere. It is important to note that this formula applies to both apparent and absolute scales and can be used in various magnitude systems.

To determine $M_{\rm DS}$, we model the spectrum of the DS as a blackbody.  Additionally, we assume that DSs behave as gray absorbers. Under these assumptions, the model star + DS depends on two free parameters: the covering factor ($\gamma$) and the effective temperature of the Dyson sphere ($\rm T_{DS}$). The covering factor $\gamma$ is defined as the normalized luminosity of the DS: \begin{equation}
\gamma = \frac{L_{\rm DS}}{L_{\star}},
\label{eq:gamma}
\end{equation} where $L_{\rm DS}$ is the luminosity of the DS and $L_{\star}$ is the luminosity of the star hosting the DS before being obscured. Under this definition, $\gamma$ can only be a positive number lower or equal to 1. In the case of an isotropically radiating star, $\gamma$ also represents the fractional solid angle of outgoing radiation intercepted by the DS (the covering factor) or the DS's completion level if we assume that the structure is nearly spherical. With all this information, we can determine the magnitude of the star when it is obscured by the DS using the following Equation: \begin{equation} M_{\star} = M_{\star,O} - 2.5\log_{10} (1 - \gamma),
\label{eq:mstar}
\end{equation} where $M_{\star,O}$ is the magnitude of the star before being obscured. In practice, we take $M_{\star,O}$ values from main-sequence stars in the Gaia-2MASS-AllWISE photometry as described below.

\begin{figure}
\centering
    \begin{subfigure}[b]{0.48\textwidth}
    \centering
    \includegraphics[width=\textwidth]{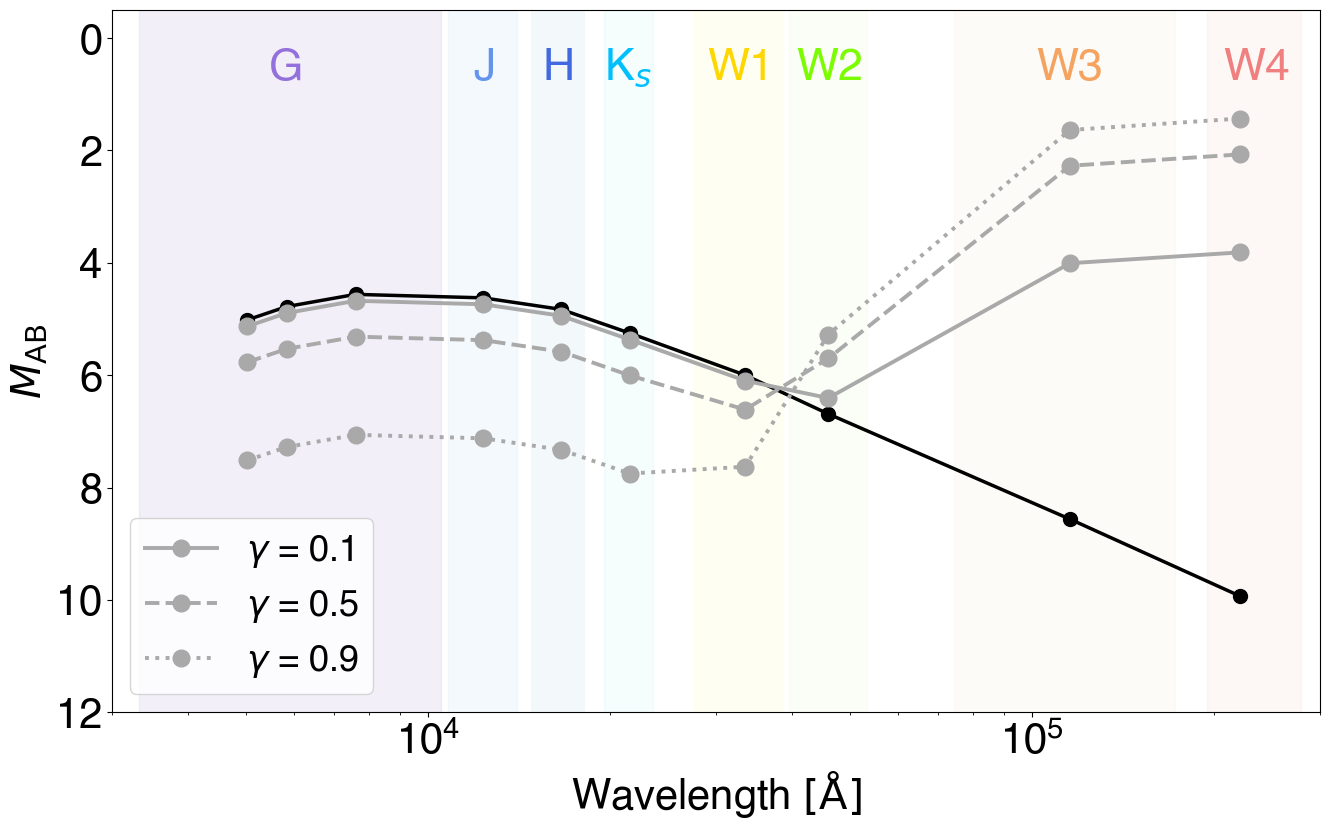}
    \end{subfigure}
    \begin{subfigure}[b]{0.48\textwidth}
    \centering
    \includegraphics[width=\textwidth]{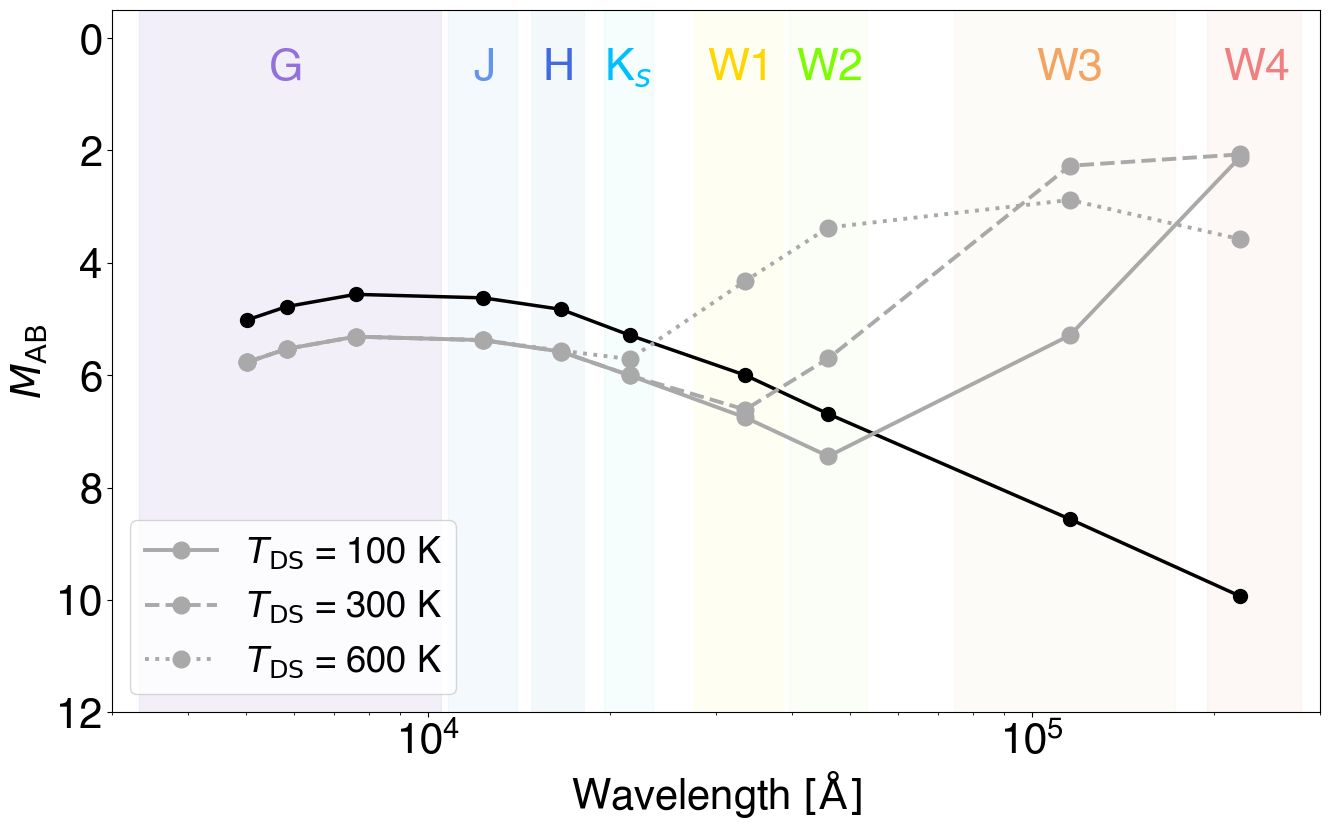}
    \end{subfigure}
    \caption{Modified photometry of a Sun-like star in the Gaia-WISE-2MASS bands due to the presence of various Dyson spheres. The unmodified absolute magnitudes of the Sun-like star ($T \rm _{eff}$ = 5777 K) are represented by solid black lines. In the top panel, the Dyson sphere models have an effective temperature of $T \rm _{DS}$ = 300 K and covering factors of 0.1, 0.5, and 0.9, depicted by solid grey, dashed, and dotted lines, respectively. In the bottom panel, the Dyson sphere models have a fixed covering factor of $\gamma$ = 0.5 and temperatures of 100, 300, and 600 K, depicted by solid grey, dashed, and dotted lines, respectively. The colored bands in the plots represent the wavelength ranges detectable by the Gaia, 2MASS, and WISE missions. It is important to note that the absolute magnitudes depicted in these plots are in the AB system.}
    \label{fig:ex1}
\end{figure}

In summary, Equations~\ref{eq:mag} and ~\ref{eq:mstar} provide a framework for understanding the changes in the magnitude of a star if it were hosting a DS. These equations describe the transformation from the original magnitude $M_{\star,O}$ to the modified magnitude $M$ when considering a Dyson sphere with a given temperature $T_{\rm DS}$ and covering factor $\gamma$. We also assume that Dyson spheres are built up slowly and uniformly everywhere, with equal covering factor ($\gamma$) in every direction, with no pieces large enough to cause stellar variability, see Section \ref{sec:variability}. An interesting feature of this model is that it is identical to optically thin blackbody debris disk models, where the covering factor $\gamma$ resembles the fractional luminosity (L$\rm _{Disk}$/L$_\star$). Figure~\ref{fig:ex1} illustrates examples of the photometry of a Sun-like star ($T_{\rm eff} = 5777$ K) hosting Dyson spheres with various parameters. In the top panel, the composite spectrum is shown for a fixed DS temperature of 300 K and covering factors of $\gamma$ = 0.1, 0.5, and 0.9, while the bottom panel displays the spectrum variations for a fixed covering factor of 0.5 and DS temperatures of 100, 300, and 600 K. The main signatures produced by a Dyson sphere include a drop in stellar flux and a boost of the flux in the mid-IR, where the mid-IR peak depends on the temperature of the Dyson sphere. The figure demonstrates how the crucial infrared information required for the identification of Dyson sphere candidates is contained within the W3 and W4 bands, as mentioned in Section~\ref{sec:init}. Consequently, we demand that all stars that undergo our analysis have detections in both W3 and W4 bands.

Although the temperature of the Dyson sphere is a free parameter, we limit our search to Dyson sphere temperatures ranging from 100 to 700 K to align with WISE's infrared detection capabilities. Additionally, we consider covering factors equal to or greater than 0.1, as this threshold ensures significant infrared excess for detection, as shown by \citet{suazo22}. In total, we generated 220,745 Dyson sphere models by simulating how the Gaia-2MASS-WISE photometry of 265 main-sequence stars would change in the presence of Dyson spheres according to the presented models. We select main-sequence stars with $M_G$ values ranging from 0 to 13.6 (stellar effective temperatures from $\sim$2,800 to 12,500 K) and ensure that these are main-sequence stars as explained in Appendix~\ref{sec:models}. We also ensure that these stars already do not possess any mid-infrared excess.

\subsection{Grid Search}
\label{sec:grid}

After generating the 220,745 models, we proceeded to compare the photometry of all remaining main-sequence stars from Section~\ref{sec:init} to these models. This involves performing a grid search to find the best-fitting model for each of the 320,000 sources. The selection of the best-fitting model for each star was based on minimizing the root mean squared error (RMSE) between the observed data and the model predictions.

Following the search for the best models, we filtered out all stars whose best model yielded an RMSE higher than 0.2 mags. This selection is quite simple and does not consider the error measured since, otherwise, it would prioritize better fits in the optical rather than in the MIR, where the information of the infrared excess lies in this work. The selection of this threshold is a free parameter. Still, we chose it to be 0.2 magnitudes to reduce the sample of potential candidates to a reasonable number that we could potentially aim to follow up with additional observations on a reasonable timescale. Additionally, the selection of this threshold is motivated by comparing our models with \citet{Vioque2020} pre-main sequence, Classical Be stars, and sources that have been proposed as candidates of these two categories based on different features (photometry, optical variability, etc), but have not yet been confimed. We assessed what root mean square error (RMSE) threshold value is reasonable by comparing our models to the photometry of the stars presented in this catalog. Since pre-main sequence stars and Classical Be stars are known to be significant sources of mid-infrared emission and, therefore, represent potential interlopers in our search. Most stars in the \citet{Vioque2020} catalog that we examined displayed an RMSE higher than 0.2 magnitudes when compared to our models, so we used this threshold as our goodness-of-fit criterion to select potential candidates. We found $\sim$11,000 sources whose best fit suffices an RMSE lower than 0.2. 

After filtering the stars based on the RMSE criterion, we proceeded to classify the remaining sources using a neural network. This classification aimed to distinguish whether the sources were located in nebular regions. Nebulae can generate features that are similar to those hypothetically produced by a Dyson sphere, hence the motivation behind developing this algorithm. 

\subsection{Image classification}
\label{sec:cnn} 

Upon selecting candidates using an RMSE as our goodness-of-fit metric, we found that young dust-obscured stars or stars otherwise associated with prominent nebulae appear as common false positives. Previous searches for infrared sources \citep[e.g.,][]{ribas2012,kennedy2012} encountered contamination issues due to the presence of foreground or nearby sources, which can cause large photocenter shifts across all WISE bands and/or an extended morphology. All these phenomena can produce photometric signatures that resemble those of our models. 
To reduce the number of interlopers in the form of young obscured stars in our sample, we developed an algorithm to classify whether stars lie or not in nebular areas based on their WISE images. This algorithm utilizes normalized W3 images as input and aims to classify stars based on whether they reside in nebular regions. The CNN architecture 
employed in this work is presented in Table~\ref{tab:cnn}, and it was developed using the PyTorch library \citep{pytorch3}.

\begin{figure}
    \centering
    \includegraphics[width=0.9\columnwidth]{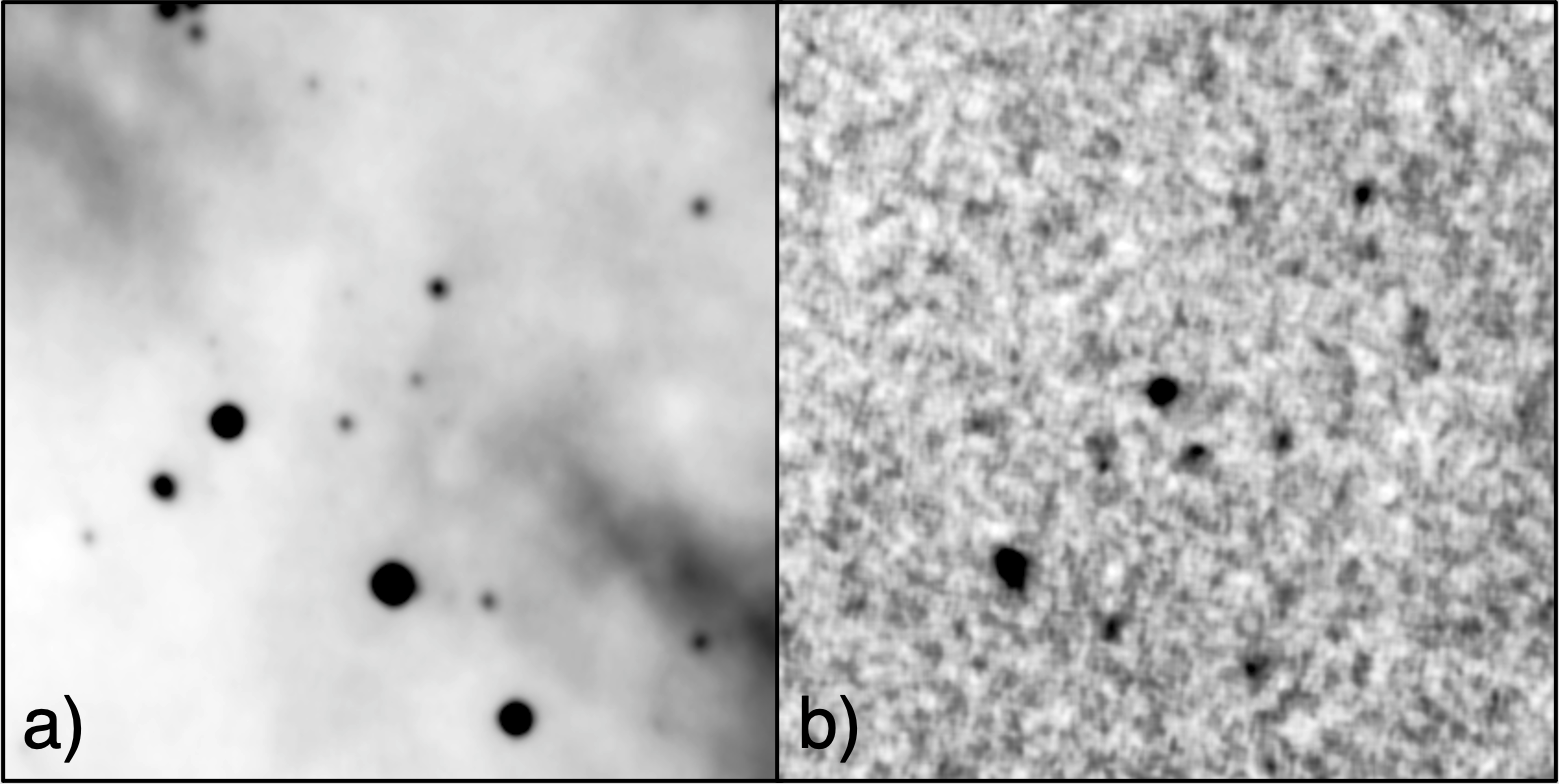}
    \caption{Two images exemplify each category's appearance: Nebular on the left-hand side panel and non-nebular on the right-hand side. Both images are normalized. Each image corresponds to a squared region in the sky with a side of 9.625 arcmin.}
    \label{fig:example_nebular}
\end{figure}

Our algorithm's input images were standardized to 420 × 420 pixels, with each pixel representing a square of side 1.375 arcsec. This corresponds to a squared image with a side of 9.625 arcmin. Then, we classified 960 images by ocular inspection, with half of them depicting images of stars embedded in nebulae and the other half representing non-nebular cases. In Figure~\ref{fig:example_nebular}, we provide examples of two images that were classified as nebular and non-nebular. We split our sample into the training, validation, and testing subsets. All subsets were built by selecting random images in our sample. Training, validation, and testing sets were randomly sampled and split into 70\%, 15\%, and 15\% of the total dataset, respectively. 

We do not include W1 nor W2 bands since dusty nebular features are typically not detectable in these bands. We also omit the W4 images since these tend to have lower quality and do not provide much extra information compared to W3. 

The specific CNN architecture used in this work is presented in Table~\ref{tab:cnn}. For the convolutional layers, the parameters shown in Table~\ref{tab:cnn} are the filter dimensions and the number of output channels. No padding was applied to any of the convolutional layers. Moreover, to all the convolutional and fully connected layers, a Rectified Linear Unit \citep[ReLU;][]{relu} activation function was applied, except for the last fully connected layer, which utilized a softmax function instead. Additionally, we "batch normalized" \citep{bn} every layer after convolution. The output of the last convolutional layer is flattened to feed the fully connected layers. We additionally applied a dropout regularization after each layer in the fully connected network \citep{dropout}. We seek the minimum of the loss function by using the Adam algorithm \citep{adam}. The network was trained using batches of 64 images.

To optimize the performance of our classifier, we conducted a hyper-parameter search by randomly sampling 79 out of the 5184 possible combinations of the parameters listed in Table~\ref{tab:hyper}. The parameters that were tuned include the learning rate, the beta parameters ($\beta_1$ and $\beta_2$) of the Adam algorithm, the dropout probability ($p$), the number of neurons in the fully connected network, and the kernel size in the convolutional matrices.

The learning rate controls the magnitude of weight updates during training, whereas the beta parameters $\beta_1$ and $\beta_2$ are decay rates used to estimate the moments of the gradient for finding the global minimum of the loss function. The dropout probability $p$ determines the probability of zeroing out a neuron in a layer to prevent overfitting. The number of neurons in the fully connected network determines the number of units in each hidden layer. It is important to note that the learning rate and beta parameters are related to the training process, while the dropout probability and the number of neurons per hidden layer are design parameters of the architecture.

\begin{table}
    \centering
    \begin{tabular}{c|c|c}
    \hline
        Layer & Layer Parameters & Output Size\\
    \hline
        Input & & 420 $\times$ 420 $\times$ 1 \\
        Convolution$\rm ^{a,b}$ & 3 $\times$ 3, 6 & 418 $\times$ 418 $\times$ 6 \\
        Max pooling & 2 $\times$ 2, Stride 2 & 209 $\times$ 209 $\times$ 6 \\
        Convolution$\rm^{a,b}$ & 3 $\times$ 3, 12 & 207 $\times$ 207 $\times$ 12 \\
        Max pooling & 2 $\times$ 2, Stride 2 & 103 $\times$ 103 $\times$ 12 \\
        Convolution$\rm^{a,b}$ & 3 $\times$ 3, 32 & 101 $\times$ 101 $\times$ 32 \\
        Max pooling & 2 $\times$ 2, Stride 2 & 50 $\times$ 50 $\times$ 32 \\
        Convolution$\rm^{a,b}$ & 3 $\times$ 3, 64 & 48 $\times$ 48 $\times$ 64 \\
        Max pooling & 2 $\times$ 2, Stride 2 & 24 $\times$ 24 $\times$ 64 \\
        Convolution$\rm^{a,b}$ & 3 $\times$ 3, 128 & 22 $\times$ 22 $\times$ 128 \\
        Flatten & & 61952 \\
        Fully Connected Network & & \\ 
        First Hidden Layer$\rm^{a,b,c}$& 61952 $\times$ 256 & 256 \\
        Second Hidden Layer$\rm^{a,b,c}$ & 256 $\times$ 256 & 256\\
        Third Hidden Layer$\rm^{a,b,c}$ & 256 $\times$ 256 & 256\\
        Softmax & 256 $\times$ 2 & 2\\
    \hline
    \end{tabular}
    \caption{Convolutional Neural Network Architecture. Batch normalization is applied to all layers with superscript $\rm ^a$. The ReLU activation function is applied to all processes with superscript $\rm ^b$. A dropout regularization was applied to all layers with superscript $\rm ^c$.}
    \label{tab:cnn}
\end{table}

\begin{table}
    \centering
    \begin{tabular}{c|c}
    \hline
        Hyperparameter & Random Search Values \\
        \hline
        Learning Rate & 10$^{-3}$, 5$\cdot10^{-4}$, 10$^{-4}$, 5$\cdot10^{-5}$ \\
        Regularization parameters ($\beta_1$,$\beta_2$) & 0, 0.3, 0.5, 0.7, 0.9, 0.99 \\
        Dropout probability $p$ & 0, 0.2, 0.4, 0.6 \\
        Number of neurons & 32, 128, 256 \\
        Kernel size & 3, 5, 7\\
        \hline
    \end{tabular}
    \caption{Hyperparameter Random Search Values}
    \label{tab:hyper}
\end{table}

We trained nine networks for each combination of hyperparameters with different initial random weights. The initial weights are sampled from the uniform distribution that PyTorch has implemented to initialize weights. Additionally, each network was trained during 35 epochs. After evaluating 79 random hyperparameter combinations, we found several combinations that yielded accuracies $\sim$93 \% on the validation set. A family of neural networks with similar characteristics and performances was identified, and the specific hyperparameters of this family and their performance are shown in Table~\ref{tab:family}. Accuracies are reported on the testing set.

\begin{table*}
    \centering
    \begin{tabular}{c|c|c|c|c|c|c|c|c}
    \hline
        Label & Learning & $\beta_1$ & $\beta_2$ & Dropout & Number & Kernel & Average & Standard \\
        Experiment &Rate & & & probability &of neurons& size & Accuracy & Deviation Accuracy \\
        \hline
        A & $5\cdot10^{-4}$ & 0.5 & 0.9 & 0.6 & 256 & 3 & 0.916 & 0.031\\
        B & $10^{-4}$ & 0.5 & 0.5 & 0.6 & 128 & 3 & 0.913 & 0.017\\
        C & $10^{-4}$ & 0.5 & 0.9 & 0.6 & 128 & 3 & 0.915 & 0.029\\
        D & $5\cdot10^{-4}$ & 0.5 & 0.9 & 0.6 & 128 & 3 & 0.908 & 0.020 \\
        E & $10^{-4}$ & 0.5 & 0.5 & 0.6 & 256 & 3 & 0.906 & 0.034\\
        F & $5\cdot10^{-4}$ & 0.5 & 0.5 & 0.6 & 256 & 3 & 0.930 & 0.016\\
    \hline
    \end{tabular}
    \caption{Best hyperparameter combination}
    \label{tab:family}
\end{table*}

\begin{figure}
    \centering \includegraphics[width=0.85\columnwidth]{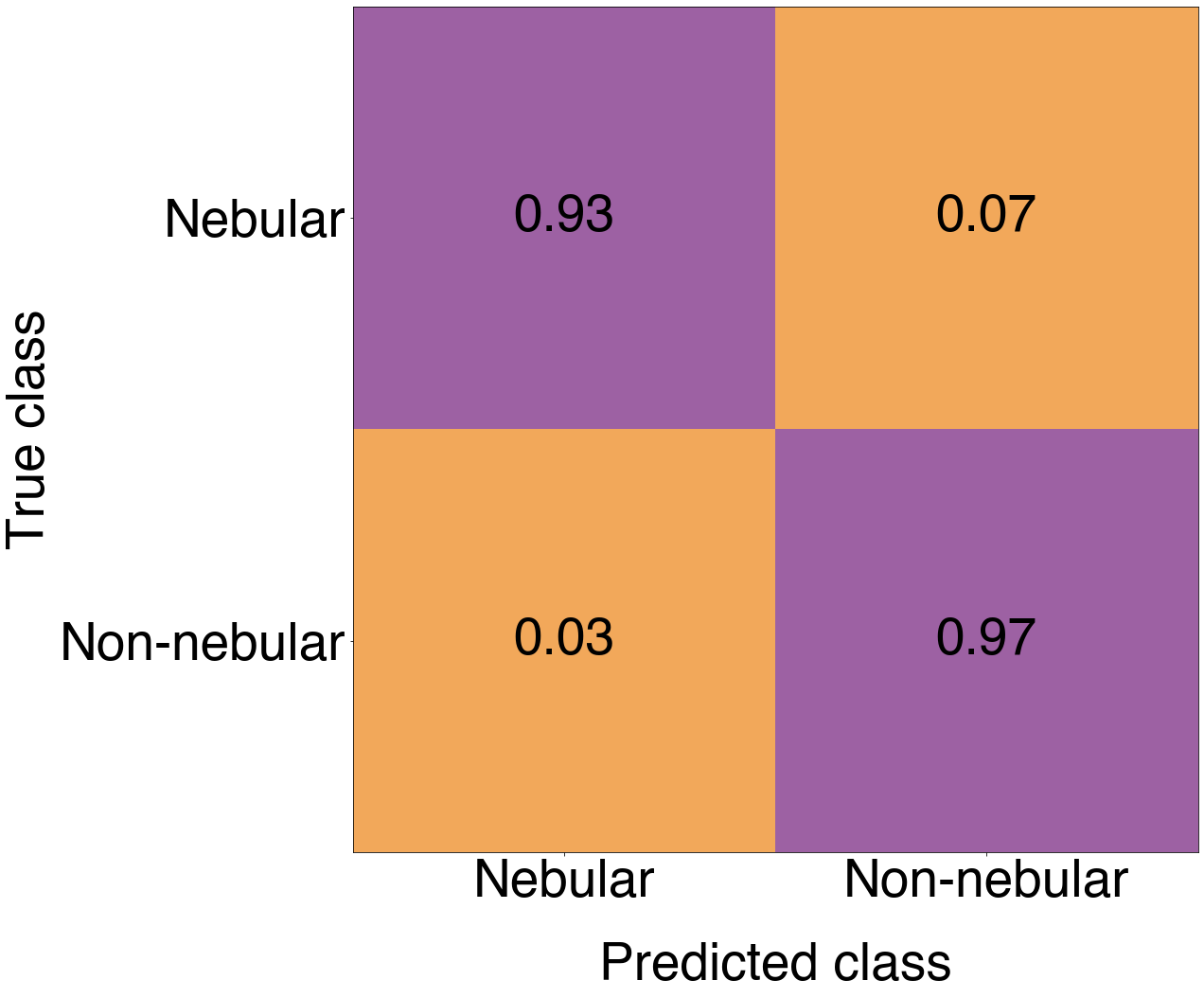}
    \caption{Normalized confusion matrix for the test set using the architecture yielding the best results. The test set contains 144 elements.}
    \label{fig:matrix}
\end{figure}

From the family of neural networks with similar performances, we selected the architecture that achieved the highest mean accuracy and the lowest standard deviation. In this case, it corresponds to experiment F in Table~\ref{tab:family}. Additionally, in Figure~\ref{fig:matrix}, we show the confusion matrix for the testing set in the best run for this architecture. The accuracy is 0.95, the recall is 0.975 on the non-nebular class, and the precision is 0.93 on the non-nebular class.

Using the trained CNN, we proceeded to classify whether stars lie or not in the nebular region. We find that 5732 sources appear as sources in non-nebular regions according to our classifier.




\subsection{Additional Analysis}
\label{sec:addiotional}

In the next subsections, we introduce additional criteria and cuts to refine further and validate our selection of Dyson sphere candidates among the sources exhibiting an infrared excess. These criteria help us rule out false positives and ensure we focus on the most promising candidates.


\subsubsection{H${\alpha}$ emission}

The emission of H${\alpha}$ photons is an important signature of young stars, particularly during strong accreting episodes. When a young protostar heats up, it ionizes the surrounding hydrogen-dominated accretion disk, which ends up emitting H${\alpha}$ photons \citep{barrado2003}. 

In Gaia DR3, the pseudo equivalent width of H${\alpha}$ is provided as one of the new products \citep{gaiadr3_apsis1, gaiadr3_apsis2}, and it becomes one of the most important parameters when weeding out interlopers. Just as optical variability is a characteristic feature of pre-main-sequence stars, the emission of H${\alpha}$ photons due to hydrogen excitation during the accretion process is another significant signature. To filter out false positives, sources with H${\alpha}$ equivalent widths lower than zero (at 3$\sigma$) are rejected, i.e., sources with H$\alpha$ in emission detected at 99.7 \% confidence.

\subsubsection{Optical variability}
\label{sec:variability}

Pre-main-sequence stars, being in the early stages of stellar evolution, can naturally emit infrared radiation due to the presence of an accretion disk surrounding the forming star. These young stars often exhibit brightness variability as a characteristic feature \citep[e.g.,][]{Joy1945,Herbst2007}. The variability can be attributed to various factors, including circumstellar obscuration events, hot spots on the star or disk, accretion bursts, and rapid structural changes in the accretion disk \citep{Cody2014}.

Gaia DR3 provides an optical variability flag among other newly added products. However, this flag is unavailable for most sources. In order to assess the optical variability of stars, we ourselves constructed the observable $G_{\rm var}$, which is defined in \citet{Vioque2020}. This observable aims to quantify the level of optical variability and has been used to classify different types of variable stars, including Herbig Ae/Be stars, TTauri stars, and Classical Be stars. The observable $G_{\rm var}$ is defined as: \begin{equation}
    G_{\rm var} = \frac{F'_{\rm G} e(F_{\rm G}) \sqrt{N_{\rm obs,G}}}{F_{\rm G} e'(F_{\rm G}) \sqrt{N'_{\rm obs,G}}},
\end{equation} where $F_G$ and $e(F_G)$ are the Gaia $G$ band flux and its uncertainty, respectively, while $N_{\rm obs,G}$ corresponds to the number of times that that source was observed in the $G$ band. The logic behind this formula relies on the fact that variable sources should have larger uncertainties compared to non-variable ones. The denominator refers to the median value of sources with similar fluxes since non-variable objects exhibit different uncertainties. \citet{Vioque2020} showed that pre-main-sequence stars exhibit a wide range of $G_{\rm var}$ that goes from $\sim$0.7 to $\sim$100. The distribution of $G_{\rm var}$ for known pre-main-sequence stars peaks at $G_{\rm var}$ $\sim$6, and it decreases toward the above-mentioned values. Here, we reject all stars exhibiting a $G_{\rm var}$ higher than two, since they are most likely to be young stars. Similarly, \citet{barber2023} developed a proxy for stellar variability and age, indicating that Gaia excess photometric uncertainties decrease linearly with $\log_{10}(\text{age})$ in Myr. However, this relation primarily applies to FGK and early M-type stars. These studies demonstrate the potential of using Gaia uncertainties and variability measures to infer the ages and variability status of stars. 

It is important to note that this check rejects potential Dyson swarms with very large absorbing elements since these in principle could generate detectable variations in the photometry of the host star. However, these variations could be mistaken for other astrophysical phenomena such as asteroseismic variations or photometric noise \citep{wright16}. It is also practical to exclude variable sources; otherwise, young stars would more easily slip through our pipeline.

\subsubsection{Astrometry}

Our search strongly relies on parallax-based distances, which can be incorrectly estimated if the single-star model fails to fit the astrometric observations. In order to assess the reliability of the distance, Gaia provides the Renormalised Unit Weight Error (RUWE), a parameter that tells us how well astrometric observations fit the astrometric solution. RUWE values tend to be close to 1.0 for well-behaved sources, while significantly higher values exceeding 1.0 may indicate non-single or problematic sources. To ensure reliable astrometry, we implemented a conservative RUWE threshold of 1.4. Sources surpassing this threshold are excluded as potential candidates to minimize objects with unreliable distance estimates. Other studies \citep[e.g.,][]{Stassun2021} have shown a significant correlation between the RUWE statistic and unresolved binary systems. Binary systems can generate warm dust through processes such as the catastrophic collision of planets \citep[e.g.,][]{weinberger2008, thompson2019}. Given that such systems might have inaccurate distances and exhibit mid-IR flux excess, the aforementioned RUWE criterion aids in rejecting sources potentially comprising binaries surrounded by warm dust, as well as those with problematic astrometry.

 \subsubsection{Extended sources}

We expect all candidates to have a shape consistent with a point source, therefore, we rule out all sources having a non-zero AllWISE $ext\_flag$.

\subsubsection{Star probability}

Gaia also classifies sources into different categories. We use one of the probability metrics Gaia DR3 provides to ensure the source is more likely to be a star. In particular, we use $classprob\_dsc\_combmod\_star > 0.9$ to consider our source candidates. We found no difference when comparing similar classification metrics. 

\subsubsection{Sources rejected so far}
Out of all the criteria outlined in Section~\ref{sec:addiotional}, the RUWE criterion refutes the largest quantity of candidates. A total of 282 sources are rejected by this criterion alone, which corresponds to roughly half of all sources rejected by any criteria in Section~\ref{sec:addiotional}. The $H_{\alpha}$ emission, the optical variability, and the extended flag criteria equally contribute to the rest of the cuts. We noticed that over 1,000 sources have negative $H_{\alpha}$ EWs. However, the uncertainties are so large that we cannot confirm $H_{\alpha}$ emission at the 3$\sigma$ level.

\subsection{SNR criterion}
\label{sec:snr}

After applying all the cuts presented in Section~\ref{sec:methods}, we ended up with 5137 sources with DS-like SEDs. Consequently, we proceeded to visually inspect some of the W3/W4 images of these candidates. This step revealed that most of them appeared to be unconvincing as secure point-like sources. In many cases, these sources appear irregular or blend with the background noise. Although WISE data reduction considers any signal with a SNR value higher than 2 as a detection, many of these detections are not reliable and fail to represent genuine infrared sources; most of the inspected images matched this pattern. Therefore, an additional cut was applied based on the SNR of these $\sim$5,000 sources. We selected sources with SNR higher than 3.5 in both the W3 and W4 bands, resulting in 368 sources. 

\begin{table}
    \centering
    \begin{tabular}{c|c}
        \hline
        Stage & Number of stars \\
        \hline
        Stars in Gaia DR3-2MASS-AllWISE & $\sim$5 $\cdot 10^6$ \\
         within 300 pc & \\
         W3/W4 detection & $\sim$3.2 $\cdot 10^5$\\
         RMSE $\leq$ 0.2 & 11243 \\
         Nebular classifier & 5732 \\
         Extra cuts & 5137 \\
         SNR W3/W4 > 3.5 & 368 \\
         Final Candidates & 7 \\
        \hline
    \end{tabular}
    \caption{Number of stars after every cut.}
    \label{tab:numbers}
\end{table}

\subsection{Visual Inspection}
\label{sec:visual}

\begin{figure*}
    \centering
    \includegraphics[width=0.8\textwidth]{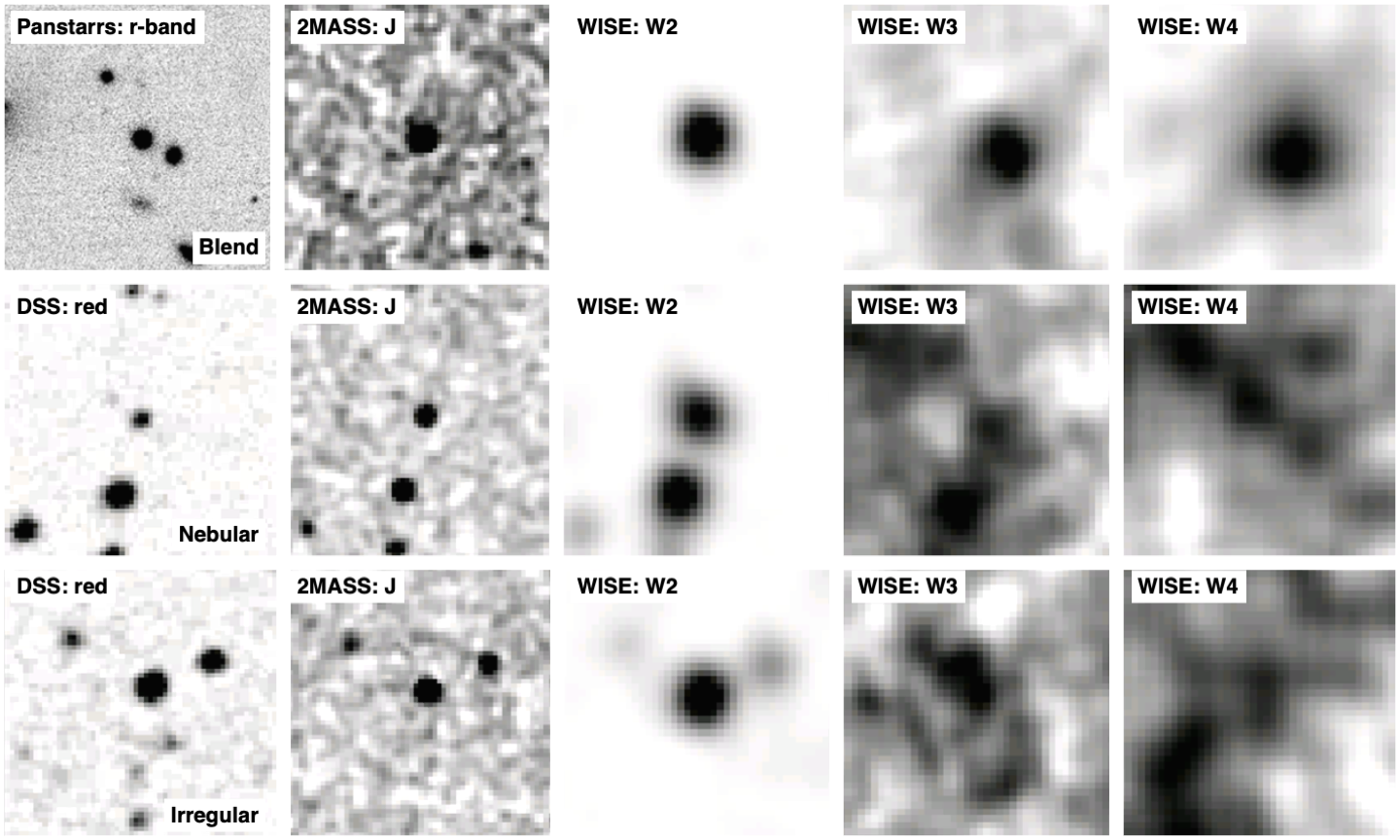}
    \caption{Examples of typical confounders in our search. The top row features a source from the blends category, the middle row a source embedded in a nebular region, and the bottom row a case from the irregular category. On these scales, the irregular and nebular cases cannot be distinguished, but the nebular nature can be established by inspecting the images at larger scales.}
    \label{fig:confounders}
\end{figure*}

After rejecting all sources with low SNRs, we conducted a second pass of visual inspections for all sources that survived the SNR cut. Visual examination of WISE images \citep[e.g.,][]{ribas2012,sgro2021} is a common technique to identify and reject unreliable sources, as not all flags or metrics provided by WISE can address issues in the data reduction. Following scrutiny of all WISE images, we categorized three types of confounders: blends, irregular structures, and nebular features. Figure~\ref{fig:confounders} illustrates the distinctions between these classes. In the top row, we showcase the 'blend case,' where a source overlaps with external sources within the aperture of the WISE bands, particularly noticeable in the W3 and W4 bands. Optical images with higher resolution facilitate the detection of blends. Even if some contaminants do not emit optical light, if an infrared source appears significantly shifted from the image center and lacks optical emission, it is considered a blend and subsequently rejected.

In the second row of Figure~\ref{fig:confounders}, we depict the "nebular" category of false positives. These cases exhibit W3 and W4 images that appear hazy and disordered, lacking a discernible source of infrared radiation at the location of the candidate. However, upon examining large-scale images spanning approximately 600 arcseconds, distinctive nebular features become evident. Some of these features resemble the example shown in Figure~\ref{fig:example_nebular}. These confounding sources are instances where our Convolutional Neural Network (Section~\ref{sec:cnn}) failed to reject these sources accurately. In the third row, we illustrate the "Irregular" category, which encompasses all sources that deviate from a point-like source in their W3 and W4 bands despite being selected based on having WISE \textit{ext\_flag} values equal to 0. In this category, the sources of irregularities in our candidates’ W3 and W4 images are unclear, and there seems to be no indication of nebulosity in their surroundings when looking at larger-scale images. Causes of irregularities could be attributed to faint nebular features, high noise, and blends, but it is challenging to pinpoint the exact cause of this phenomenon. Most sources rejected in the SNR criterion had WISE images that would have fallen into this category.

\begin{figure}
    \centering
    \includegraphics[width=\columnwidth]{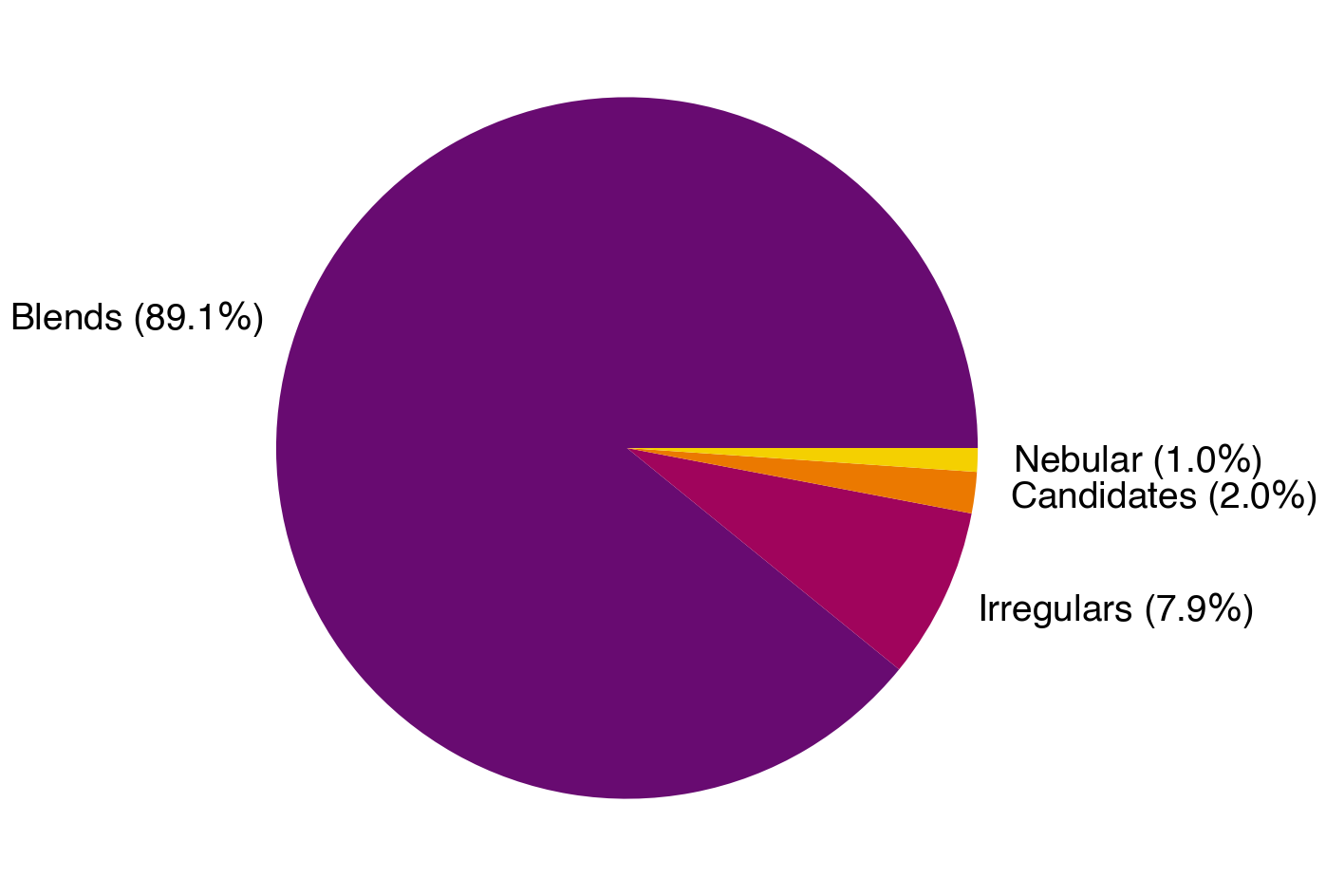}
    \caption{Pie chart illustrating the cause of infrared radiation according to our extra inspection.}
    \label{fig:pie}
\end{figure}

Among the 368 sources that survived the last cut, we identified 328 (89.1$\%$) sources as blends, 29 (7.9$\%$) as irregulars, and 4 as nebular (1.0 $\%$). After this analysis, a total of 7 (2.0 $\%$) sources were identified as potential candidates that appear to be free of conspicuous problems. The visual inspection results are summarized in Figure~\ref{fig:pie}. Many blends were identified thanks to the inspection of optical images, so we double-checked that our seven final sources were free of contaminants by examining Pan-STARRS1 DR1 \citep{panstars} and Sky Mapper DR2 \citep{skymapper} images to account for both hemispheres. None of these seven sources showed any indication of contamination.

Finally, for the seven sources identified as potential candidates, we conducted a search for nearby X-ray sources. X-rays are a powerful tool for tracing star-forming regions in the sky \citep[e.g.,][]{scortino2022}, suggesting our candidates could be young stars if X-ray sources associated with star formation were present in their vicinity. After searching the XMM-Newton science archive, we found no evidence of X-ray sources in the neighborhood of our candidates that could be attributed to star formation. In one instance, there is an X-ray source approximately 14 arcminutes from a candidate; however, this source is confirmed to be a Seyfert galaxy.

\section{Results}
\label{sec:results}

In Table~\ref{tab:candidates}, we summarize all candidates. Our visual inspection indicates that these sources are actual sources of infrared radiation that are not subject to any obvious contamination. Given the limited number of candidates, we revised our model fitting using a more refined grid compared to the one employed in Section~\ref{sec:grid}. This time, we compared our data to 6,216,900 models, encompassing 391 Dyson sphere effective temperatures ranging from 10 to 400 K and 60 covering factors ranging from $10^{-4}$ to 0.4. Table~\ref{tab:candidates} presents the updated Dyson sphere temperature estimates and covering factors. 

\begin{table*}
    \centering
    \begin{tabular}{c|c|c|c|c|c|c|c|c|c|c}
        \hline
        Label & Gaia DR3 ID & Distance$\rm ^{a}$ [pc] & $m \rm _G$$\rm ^{b}$ & $G \rm _{var}$$\rm ^{c}$ & RUWE$\rm ^{b}$ & $T\rm _{eff}$$\rm ^{b}$ [K] & EWH$\alpha$$\rm ^{b}$ [nm] & $T_{\rm DS}$$\rm ^{c}$ [K] & $\gamma$$\rm ^{c}$ & S/N$\rm ^{d}$ (W3/W4) \\
        \hline
        A & 3496509309189181184 & 142.9 $\pm$ 1.0 & 15.99 & 1.03 & 1.03 & - & 0.248 $\pm$ 0.076 &  138 $\pm$ 6 &  0.08 $\pm$ 0.01 & 22.5 / 16.6 \\
        B & 4843191593270342656 & 211.6 $\pm$ 3.5 & 17.71 & 0.94 & 1.06 & 3574 & - &  275 $\pm$ 40 & 0.06 $\pm$ 0.008 & 13.9 / 3.8 \\ 
        C & 4649396037451459712 & 219.4 $\pm$ 6.2 & 18.39 & 0.90 & 1.21 & 3238 & - & 187 $\pm$ 16 & 0.14 $\pm$ 0.016  & 10.5 / 5.0 \\
        D & 2660349163149053824 & 211.5 $\pm$ 5.8 & 17.66 & 0.97 & 0.96 & 3473 & - & 178 $\pm$ 20 & 0.16 $\pm$ 0.03 & 10.4 / 4.8 \\
        E & 3190232820489766656 & 274.7 $\pm$ 6.1  & 17.00 & 0.90 & 1.05 & 3556 & 0.049 $\pm$ 0.100 & 180 $\pm$ 26 & 0.08 $\pm$ 0.02 & 10.3 / 3.6 \\
        F & 2956570141274256512 & 265.0 $\pm$ 2.6 & 16.32 & 0.93 & 1.01 & 3674 & 0.020 $\pm$ 0.068 & 137 $\pm$ 16 & 0.03 $\pm$ 0.008 & 5.7 / 4.5 \\
        G & 2644370304260053376 & 249.9 $\pm$ 3.7 & 16.48 & 0.99 & 1.01 & 3480 & 0.024 $\pm$ 0.097 & 100 $\pm$ 9 & 0.13 $\pm$ 0.02 & 5.0 / 3.5 \\
        \hline
    \end{tabular}
    \caption{Dyson sphere candidates. All sources are clear mid-infrared emitters with no clear contaminators or signatures that indicate an obvious mid-infrared origin. We present data derived from
    $\rm ^{a}$ Gaia EDR3 \citet{Bailer-Jones21}. $\rm ^{b}$ Gaia DR3. $\rm ^{c}$ This work. $\rm ^{d}$ AllWISE \citet{cutri}.}
    \label{tab:candidates}
\end{table*}

\begin{figure*}
    \centering
    \includegraphics[width=\textwidth]{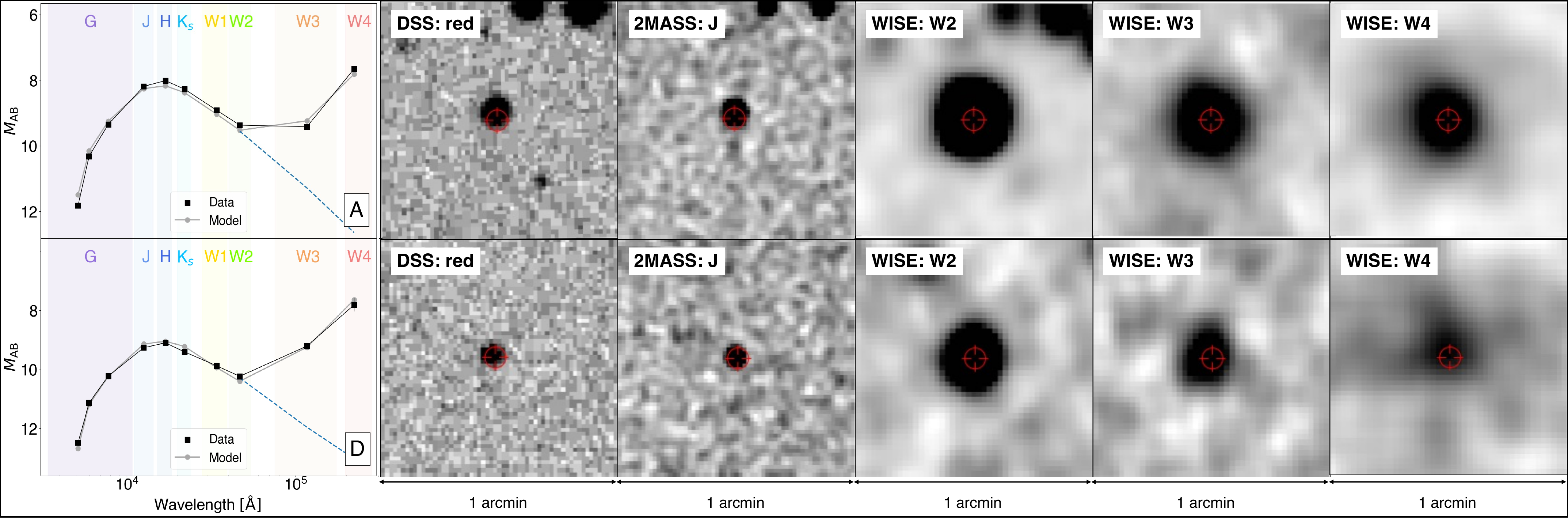}
    \caption{SEDs of our two Dyson spheres candidates and their photometric images. The SED panels include the model and data, with the dashed blue lines indicating the model without considering the emission in the infrared from the Dyson sphere and the solid black line indicating the model that includes the infrared flux from the Dyson sphere. Photometric images encompass one arcmin. All images are centered in the position of the candidates, according to Gaia DR3. All sources are clear mid-infrared emitters with no clear contaminators or signatures that indicate an obvious mid-infrared origin. The red circle marks the location of the star according to Gaia DR3.}
    \label{fig:sed}
\end{figure*}

While examining the pseudo-equivalent width of $H_\alpha$, we observed that some candidates exhibit too high uncertainties. Hence, there remains a possibility that some of these sources are indeed H$\alpha$ emitters, which would reveal the early stellar evolutionary stage and explain their infrared radiation. Figure~\ref{fig:sed} showcases the SEDs and photometric images of two of the seven candidates, while Table~\ref{tab:candidates} provides additional information used in our further analysis (Section~\ref{sec:addiotional}). In the examples depicted in Figure~\ref{fig:sed}, clear W3/W4 images indicate a distinct source of mid-infrared radiation in both bands. Candidate A notably displays a considerable shift between DSS, 2MASS, and the WISE images, which is attributed to its relatively high proper motion. According to Gaia DR3, this star has a proper motion of $-88.7$ mas/yr in Declination.

\subsection{Potential contamination}
\label{sec:contamination}

In this search, we encountered various sources of false positives, as detailed in previous sections. As highlighted in earlier studies \citep[e.g.,][]{kennedy2012,krivov2013,gaspar2014}, Galactic background contamination and chance alignments with extragalactic sources can induce a false infrared excess at the location of a star. In the context of investigating WISE infrared stars within the Kepler field-of-view, \citet{kennedy2012} found that the Improved Processing of the IRAS Survey (IRIS: \citet{iris}) offers valuable insights into potential background contamination. They identified that sources within regions where the 100 $\mu m$ background level exceeded 5 MJy/sr were susceptible to galactic contamination. To assess whether our Dyson sphere candidates were prone to such contamination, we utilized the IRIS maps at 100 $\mu m$ to evaluate the background level of our sources. Table~\ref{tab:iris} summarizes these values, all of which fall below the threshold suggested by \citet{kennedy2012}. This result stems from our procedure of filtering out all stars embedded in nebular regions, thereby naturally eliminating sources located in regions where the Galactic background level affects the WISE photometry of stars.

\begin{table}
    \centering
    \begin{tabular}{c|c|c}
        \hline
        Label & Gaia DR3 ID & IRIS 100 $\rm \mu$m background\\
        &&  level [MJy/sr] \\
        \hline
        A & 3496509309189181184 & 4.77 \\
        B & 4843191593270342656 & 1.34 \\ 
        C & 4649396037451459712 & 4.75 \\
        D & 2660349163149053824 & 4.17 \\
        E & 3190232820489766656 & 4.45 \\
        F & 2956570141274256512 & 1.78 \\
        G & 2644370304260053376 & 2.92 \\
        \hline
    \end{tabular}
    \caption{Dyson sphere candidates and their 100 $\mu$m background level.}
    \label{tab:iris}
\end{table}

In addition to background contamination, chance alignments with bright sources in the infrared but obscured in the optical present another potential contamination source. \citet{kennedy2012} estimated the likelihood of such alignments by comparing galaxy counts with the counts of their infrared excess sources. As our Dyson sphere candidates are limited to only 7, we adopted a method akin to that used by \citet{theissen2017}. In their study, which investigates the presence of warm dust around M dwarfs, \citet{theissen2017} reanalyzed the source extraction of their targets to determine offsets among their W1, W2, and W3 images. These offsets were then compared to the inherent offset of stationary objects like quasars. Quasars serve as valuable indicators of the WISE instrument's astrometric precision as they remain stationary in the sky. \citet{theissen2017} focused solely on isolated quasars (with no other sources within 6 arcseconds), with W3 signal-to-noise ratios (SNRs) between 3 and 5, at galactic latitudes higher than 77 degrees. They noted that the offset distributions resembled those of their disk candidate stars, both exhibiting Gaussian distributions. One distribution reflected the Right Ascension offset between the W1 and W3 positions ($\mu = 0''.08$, $\sigma = 5''.00$) and another for the Declination offset between the W1 and W3 positions ($\mu = -0''.21$ a, $\sigma = 5''.48$).

In order to assess the probability of chance alignments with extragalactic sources, we adopted a similar approach and re-conducted the source extraction to determine the offset between W1, W2, and W3 images. Initially, we obtained unWISE images of our candidates. unWISE \citep{unwise} provides a collection of WISE co-added images that remain unblurred, preserving their intrinsic resolution. Subsequently, we performed a revised source extraction using the sep software \citep{sep}, a Python implementation that encompasses the core algorithms of Source Extractor (SEXtractor: \citet{sex}).

Table~\ref{tab:offset} summarizes the offsets between the positions of the extracted sources in different filters. It is noteworthy that for the W1-W2 offset, both in RA and DEC, the discrepancy is minimal and falls within the range obtained by \citet{theissen2017} in both RA and DEC. Similarly, the offsets between the W1 and W3 bands also align with the distribution, except for candidate G, which appears suspicious and warrants careful consideration. However, the current dataset lacks definitive evidence to either confirm or dismiss this candidate.

\begin{table}
    \centering
    \begin{tabular}{c|c|c|c}
        \hline
        Label & Gaia DR3 ID & W1/W2  &  W1/W3 \\
        && offset [arcsec] & offset [arcsec] \\
        && RA/DEC & RA/DEC\\
        \hline
        A & 3496509309189181184 & -0.25 / -0.01 & -0.03 / 0.33 \\
        B & 4843191593270342656 & 0.40 / 0.31 & 3.21 / 0.06 \\ 
        C & 4649396037451459712 & 0.25 / -0.32 & 1.52 / -3.68\\
        D & 2660349163149053824 & -0.31 / -0.12 & 0.60 / -0.09 \\
        E & 3190232820489766656 & -0.09 / 0.48 & -1.15 / -0.38  \\
        F & 2956570141274256512 & 0.03 / 0.10 & -1.04 / 0.79 \\
        G & 2644370304260053376 & 0.24 / 0.00 & 5.59 / 0.64 \\
        \hline
    \end{tabular}
    \caption{Offset in the photocenter of our sources in different WISE bands.}
    \label{tab:offset}
\end{table}

\section{Discussion}
\label{sec:discussion}

We conducted a comprehensive search for sources exhibiting spectral energy distributions (SEDs) compatible with stars hosting partial Dyson spheres. The last search of this kind was carried out by \citet{Carrigan09}, who only looked for complete Dyson spheres ($\gamma$ = 1) using IRAS data. We analyzed a significantly larger sample of approximately 320,000 sources from the Gaia DR3-2MASS-AllWISE dataset with W3/W4 detection, which is nearly 30 times larger than Carrigan's sample. As a result, we identified seven sources displaying mid-infrared flux excess of uncertain origin. Various processes involving circumstellar material surrounding a star, such as binary interactions, pre-main sequence stars, and warm debris disks, can contribute to the observed mid-infrared excess \citep[e.g.][]{cotten2016}. \citet{kennedy2013} estimates the occurrence rate of warm, bright dust. The occurrence rate is 1 over 100 for very young sources, whereas it becomes 1 over 10,000 for old systems (> 1 Gyr). However, the results of our variability check suggest that our sources are not young stars. If our candidates were young stars, that could explain the infrared excess and would match the more likely occurrence rate. Nevertheless, it is worth noting that although uncommon, literature has documented the existence of pre-main sequence stars with low $G\rm _{var}$ values \citep[e.g.,][]{Vioque2020}. On the other hand, our astrometric checks, which heavily rely on the RUWE parameter, indicate that the single-star astrometric solution is applicable to our sources. Despite the fact that we chose conservative thresholds for the $G\rm _{var}$ and RUWE parameters (2 and 1.4, respectively), our candidates have values that lie far below the thresholds chosen. The $G\rm _{var}$ and RUWE values are typically around unity.

\begin{figure}
    \centering
    \includegraphics[width=\columnwidth]{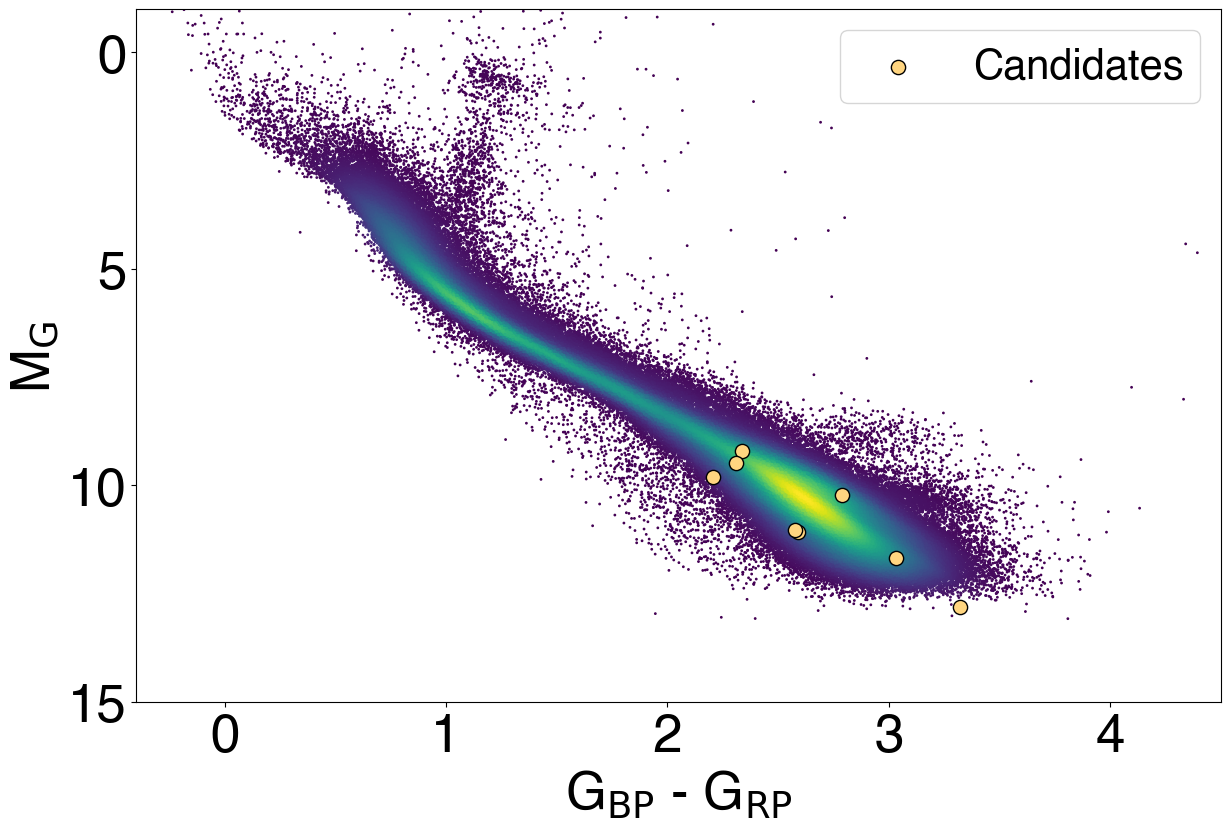}
    \caption{Color-magnitude diagram displaying the distribution of our candidates in orange circles. Colored dots represent Gaia DR3 stars within 300 pc. The color scale represents the relative density of stars.}
    \label{fig:cmd}
\end{figure}

The presence of warm debris disks surrounding our candidates remains a plausible explanation for the infrared excess of our sources. However, our candidates seem to be M-type main sequence stars, given their stellar parameters and location in the Hertzsprung-Russell diagram as Figure~\ref{fig:cmd} illustrates. However, M-dwarf debris disks are very rare objects, and up to date, only a reduced number has been confirmed \citep[e.g.,][]{luppe20,cronin2022,cronin2023}. Multiple explanations have been invoked to explain the dearth of debris disks around M dwarfs, including detection biases \citep{heng13,kennedy2018} and age biases \citep{riaz06,avenhaus2012}. Additionally, studies have suggested that the physical processes governing debris disk evolution around M dwarfs may differ significantly from those observed in solar-type stars \citep{plavchan2005}. However, the temperature and the fractional infrared luminosity ($f$ = L$_{\rm IR}$/L$_{\star}$) of our candidates are different from those of typical debris disks, which tend to be cold (10 - 100 K) and to have low fractional luminosities ($f$ $<$ 0.01). These high fractional luminosities (if we consider $f$ $=$ $\gamma$) is a feature more compatible with young disks compared to those of ordinary debris disks \citep{wyatt2008}, but the lack of variability seems to be inconsistent with the young-star scenario. On the other hand, Extreme Debris Disks (EDD) \citep{balog09}, are examples of mid-infrared sources with high fractional luminosities ($f$ $>$ 0.01) that have higher temperatures compared to that of standard debris disks \citep{moor21}. Nevertheless, these sources have never been observed in connection with M dwarfs. Are our candidates' strange young stars whose flux does not vary with time? Are these stars M-dwarf debris disks with an extreme fractional luminosity? Or something completely different? 

Several searches for infrared sources \citep[e.g.,][]{kennedy2012,ribas2012,cotten2016,theissen2017} have faced challenges in confirming authentic infrared sources. \citet{kennedy2012} demonstrated a strong correlation between the 100 $\mu m$ background level from IRIS maps and contamination, setting a 5 MJy/sr threshold to circumvent spurious infrared sources. Fortunately, this was not a concern for our candidates as we utilized a CNN algorithm, leveraging W3 images to eliminate sources within nebular regions, typically linked to high levels of far-infrared radiation near the galactic plane. Detecting infrared sources also raises concerns about potential chance alignments with infrared galaxies, leading to significant WISE photometry contamination. Various methods exist to assess the likelihood of encountering such occurrences. \citet{kennedy2012} compares extragalactic counts to their source counts, while \citet{theissen2017} re-extracts sources to compare their W1/W2/W3 positions. Following the \citet{kennedy2012} idea, we determine the contamination rate due to background galaxies that could alternatively explain the mid-infrared properties of our candidates. The contamination rate mainly depends on the number of galaxies in the sky per unit of solid angle that can produce a specific signature. In order to determine that value, we compute the number of galaxies with the following properties: W3/W4 detection with signal-to-noise ratios higher or equal to 3.5, $\textit{ext\_flg} = 0$, $W1 -W3/W4 > 1.2$ as a color cut to ensure stars are removed \citep{jarrett2011}, and 2.84  $ < W3 - W4 < $ 3.25 to ensure galaxies with a color compatible with that of our Dyson sphere models for our candidates. The total number of galactic sources per unit of solid angle is $\sim$15,000 objects/sr, which yields a contamination rate of 1.1$\cdot 10^{-5}$ if we consider a target area of 33 arcsec$^{2}$ (3.25 arcsec of radius). Notice that this contamination rate cannot be applied to the initial sample of $\sim 5 \cdot 10^6$ since that number does not consider W3/W4 detection with signal-to-noise ratios higher or equal to 3.5. Instead, we must use it on the sample of stars with W3/W4 detection and SNR $\geq 3.5$ in these bands, corresponding to $\sim$ 200,000 sources, which ultimately leads to $\sim$2 contaminated sources with the above-listed properties.

Additionally, the offsets between positions within different bands can be used as a tracer of confusion. The offset of a source within all the WISE bands should be small, given their similar PSF FWHMs (6''.1, 6''.4, and 6''.5, respectively), and WISE astrometric precision of 0''.5\footnote{https://wise2.ipac.caltech.edu/docs/release/allsky/expsup/sec6\_4.html}. In our analysis of sources, we observed no significant offset between the W1 and W2 bands. However, when examining the W1 and W3 bands, we noticed a slightly larger offset for some sources. This aligns with the offset distribution reported by \citet{theissen2017}, consistent with the offset distribution of quasars. However, candidate G exhibited a higher RA offset than expected. Although this analysis does not indicate a significant shift for six of our candidates, the possibility of perfect alignments cannot be ruled out. Therefore, each source should be approached with caution, and the potential for such alignments should not be dismissed. It is important to note that the shift observed in the seventh object might be attributed to WISE confusion, as the contamination rate suggests. WISE confusion is quite common \citep[e.g.][]{Dennihy2020} and often unavoidable, with studies indicating that it could account for as many as 70\% of false positives regarding infrared excesses around main-sequence stars \citep{silverberg2018}. 

Upon examining the color-magnitude diagram depicted in Figure~\ref{fig:cmd} alongside our candidates, it is evident that our sample predominantly comprises M dwarfs. However, our candidates deviate from the core of the M dwarf distributions, residing toward the peripheries. The rightward edge aligns more closely with young stars progressing toward the main sequence, while the leftward edge corresponds to the optical dimming anticipated by our models, which can resemble subdwarf stars.

Additional analyses are definitely necessary to unveil the true nature of these sources. Optical spectroscopy has shown to be valuable when refuting false debris disk M dwarf candidates \citep[e.g.][]{murphy2018}, and we believe it could help us constrain different features of our sources. H$\alpha$ is typically used to find out whether a star is in a young accreting stage or not. Even though chromospheric activity in M dwarfs can lead to H$\alpha$ emission, the equivalent width (EW) of the said line can be used to distinguish accretors from just chromospheric emission \citep{barrado2003}. In the latter case, the line can be used to determine several M-dwarf characteristics, such as age, stellar rotation, and magnetic activity. Additionally, the intensity of H$\alpha$ in the case of chromospheric activity is a spectral type-dependant feature \citep[e.g.,][]{lepine2013}.

Moreover, gyrochronology can help give us more insight into the ages of our candidates by using stellar rotation as an independent proxy of age since late-type stars' rotation slows down as they age \citep[e.g.,][]{kawaler1989,Barnes2003,barnes2007,meibom2015}.




\section{Conclusions}
\label{sec:conclusions}

After analyzing the optical/NIR/MIR photometry of $\sim$5 $\cdot$ 10$^6$ sources, we found 7 apparent M dwarfs exhibiting an infrared excess of unclear nature that is compatible with our Dyson sphere models. We modeled Dyson spheres with temperatures ranging from 100 to 700 K and covering factors from 0.1 to 0.9. There are several natural explanations for the infrared excess in literature, but none of them clearly explains such a phenomenon in the candidates, especially given that all are M dwarfs.


We argue that follow-up spectroscopy would help us unveil the nature of these sources. In particular, analyzing the spectral region around H$\alpha$ can help us ultimately discard or verify the presence of young disks by analyzing the potential H$\alpha$ emission. Spectroscopy in the MIR region would be very valuable when determining whether the emission corresponds to a single blackbody, as we assumed in our models. Additionally, spectroscopy can help us determine the real spectral type of our candidates and ultimately reject the presence of confounders.

We would like to stress that although our candidates display properties consistent with partial Dyson spheres, it is definitely premature to presume that the mid-infrared presented in these sources originated from them. The MIR data quality for these objects is typically quite low, and additional data is required to determine their nature.

\section*{Acknowledgements}

The authors first would like to thank the anonymous referees for their extremely helpful comments that greatly improved the quality of this study. MS acknowledges funding from the Royal Swedish Academy of Sciences. MS and EZ acknowledge funding from the Magnus Bergvall foundation. EZ and CN acknowledge a sabbatical fellowship from AI4Research at Uppsala University. EZ and SM would like to acknowledge financial assistance through the SPARC project no P39 sponsored by the Ministry of Education, Govt. of India. EZ has also benefited from a sabbatical at the Swedish Collegium for Advanced Study. This work has made use of data from the European Space Agency (ESA) mission {\it Gaia} (\url{https://www.cosmos.esa.int/gaia}), processed by the {\it Gaia} Data Processing and Analysis Consortium (DPAC, \url{https://www.cosmos.esa.int/web/gaia/dpac/consortium}). Funding for the DPAC has been provided by national institutions, in particular the institutions participating in the {\it Gaia} Multilateral Agreement. This publication makes use of data products from the Two Micron All Sky Survey, which is a joint project of the University of Massachusetts and the Infrared Processing and Analysis Center/California Institute of Technology, funded by the National Aeronautics and Space Administration and the National Science Foundation. This publication makes use of data products from the Wide-field Infrared Survey Explorer, which is a joint project of the University of California, Los Angeles, and the Jet Propulsion Laboratory/California Institute of Technology, and NEOWISE, which is a project of the Jet Propulsion Laboratory/California Institute of Technology. WISE and NEOWISE are funded by the National Aeronautics and Space Administration. The national facility capability for SkyMapper has been funded through ARC LIEF grant LE130100104 from the Australian Research Council, awarded to the University of Sydney, the Australian National University, Swinburne University of Technology, the University of Queensland, the University of Western Australia, the University of Melbourne, Curtin University of Technology, Monash University and the Australian Astronomical Observatory. SkyMapper is owned and operated by The Australian National University's Research School of Astronomy and Astrophysics. The survey data were processed and provided by the SkyMapper Team at ANU. The SkyMapper node of the All-Sky Virtual Observatory (ASVO) is hosted at the National Computational Infrastructure (NCI). Development and support of the SkyMapper node of the ASVO has been funded in part by Astronomy Australia Limited (AAL) and the Australian Government through the Commonwealth's Education Investment Fund (EIF) and National Collaborative Research Infrastructure Strategy (NCRIS), particularly the National eResearch Collaboration Tools and Resources (NeCTAR) and the Australian National Data Service Projects (ANDS). The Pan-STARRS1 Surveys (PS1) and the PS1 public science archive have been made possible through contributions by the Institute for Astronomy, the University of Hawaii, the Pan-STARRS Project Office, the Max-Planck Society and its participating institutes, the Max Planck Institute for Astronomy, Heidelberg and the Max Planck Institute for Extraterrestrial Physics, Garching, The Johns Hopkins University, Durham University, the University of Edinburgh, the Queen's University Belfast, the Harvard-Smithsonian Center for Astrophysics, the Las Cumbres Observatory Global Telescope Network Incorporated, the National Central University of Taiwan, the Space Telescope Science Institute, the National Aeronautics and Space Administration under Grant No. NNX08AR22G issued through the Planetary Science Division of the NASA Science Mission Directorate, the National Science Foundation Grant No. AST–1238877, the University of Maryland, Eotvos Lorand University (ELTE), the Los Alamos National Laboratory, and the Gordon and Betty Moore Foundation. The Digitized Sky Survey was produced at the Space Telescope Science Institute under U.S. Government grant NAG W–2166. The images of these surveys are based on photographic data obtained using the Oschin Schmidt Telescope on Palomar Mountain and the UK Schmidt Telescope. The plates were processed into the present compressed digital form with the permission of these institutions. We made use of observations obtained with XMM-Newton, an ESA science mission with instruments and contributions directly funded by ESA Member States and NASA

\section*{Data Availability}

Most of the underlying data used in this work is in the public domain (Gaia, 2MASS, WISE). Models underlying this article are available upon request.



\bibliographystyle{mnras}
\bibliography{mnras_template} 




\appendix

\section{Selection of stars}
\label{sec:models}

In order to construct our set of Dyson sphere models, we start from a set of observed stars with available Gaia DR3-2MASS-AllWISE photometry and bolometric luminosities. We select a sample of 265 main-sequence stars within 100 pc of the Sun. To ensure the selection of main-sequence stars, we apply the same criteria as \citet{suazo22}, which exclude red giants, white dwarfs, and sources with high astrometric excess noise. The filtering process is defined by Equation~\ref{eq:postms} for removing red giants and Equation~\ref{eq:wd} for eliminating white dwarfs and sources with high astrometric excess noise. In these equations, $M_{\rm G}$ represents the absolute magnitude of the star in the $G$ band, and $G_\mathrm{BP} - G_\mathrm{RP}$ denotes its color, both measured in the Vega system.

\begin{equation}
    M_{G}\text{ }<\text{ }4 \text{ and } M_G < 7\cdot(G_\mathrm{BP} - G_\mathrm{RP}) - 3,
    \label{eq:postms}
\end{equation}

\begin{equation}
    M_G > 3\cdot(G_\mathrm{BP} - G_\mathrm{RP}) + 5.
    \label{eq:wd}
\end{equation}

In addition to the cuts ensuring only main-sequence stars, we consider only stars with flux measurements available in all relevant bands and a Renormalised Unit Weight Error (RUWE) below 1.4, ensuring that the astrometric solution is of high quality. Additionally, we include stars with FLAME luminosity estimations \citep{gaiadr3_apsis1, gaiadr3_apsis2}, which are necessary for our models. We exclude stars with contamination in their WISE photometry and stars already exhibiting a mid-infrared (MIR) excess. To ensure a diverse sample, we restrict our selection to stars with absolute magnitudes ($M_{\rm G}$) ranging from 0 to 13.6, corresponding to zero-age main sequence masses between approximately 0.15 and 3.5 solar masses ($\rm M_{\odot}$). Outside of this range, no stars meet the aforementioned criteria. We also selected the sample to homogeneously distribute the number of stars in the magnitudes range.

Additionally, to ensure our photometric measurements' accuracy, we considered the saturation limits for the WISE bands when selecting the main-sequence stars. Sources brighter than 8.1, 6.7, 3.8, and $-0.4$ mag (Vega) in W1, W2, W3, and W4, respectively, are known to be saturated, resulting in overestimated fluxes. To mitigate this effect, we applied the W2 correction proposed by \citet{cotten2016} specifically for the W2 band. However, our analysis found no significant difference when considering corrected and uncorrected W2 fluxes. This is primarily because our sources are located in the unsaturated regime, where the flux measurements are reliable without the need for correction. 

We applied Dyson sphere models with temperatures ranging from 100 to 700 K and covering factors between 0.1 and 0.9 for each selected star. Since we have 265 stars, 17 covering factors, and 49 temperatures, we end up with 220,745 models, 833 for each star (17 covering factors and 49 temperatures). Please see Figure~\ref{fig:ex1} for an example of how the model parameters alter the SED of a Sun-like star.


\bsp	
\label{lastpage}
\end{document}